\def\year{2018}\relax
\documentclass[letterpaper]{article} 
\usepackage{aaai18}  
\usepackage{times}  
\usepackage{helvet}  
\usepackage{courier}  
\usepackage{url}  
\usepackage{graphicx}  
\usepackage{graphicx}
\usepackage{subfigure}
\usepackage{caption}
\graphicspath{{/home/amit/Downloads/AuthorKit18/AuthorKit18/LaTeX}}
\usepackage{amssymb}
\usepackage{amsmath}
\usepackage{multirow}
\usepackage{array}
\usepackage{amsmath}
\usepackage{relsize}
\usepackage{epstopdf}
\usepackage{amsthm}
\usepackage{algorithm,algorithmic}
\usepackage[english]{babel}
\usepackage{csquotes}
\usepackage{graphicx}
\usepackage{lipsum}
\usepackage{titling}
\usepackage{graphicx}
\usepackage{fancyhdr}

\fancypagestyle{firststyle}
{
	\fancyhf{Accepted as a Workshop paper at WSDM 2018}
	\chead{}
	\rhead{}
	\lfoot{}
	\rfoot{}
	\cfoot{}
}

\begin{document}
	
%
\title{A Datamining Approach for Emotions Extraction and Discovering Cricketers Performance from Stadium to Sensex{}}

\author{Amit Agarwal, Brijraj Singh, Jatin Bedi, Durga Toshniwal\\
	Indian Institute of Technology Roorkee\\
	Department of Computer Science \& Engineering\\
	Roorkee,
	Uttarakhand India\\
}
\date{\vspace{-0.1ex}}
\maketitle

\begin{abstract}
Microblogging sites are the direct platform for the users to express their views. It has been observed from previous studies that people are viable to flaunt their emotions for events (eg. natural catastrophes, sports, academics etc.), for persons (actor/actress, sports person, scientist) and for the places they visit.
In this study we focused on a sport event, particularly the cricket tournament and collected the emotions of the fans for their favorite players using their tweets. Further, we acquired the stock market performance of the brands which are either endorsing the players or sponsoring the match in the tournament. It has been observed that performance of the player triggers the users to flourish their emotions over social media therefore, we observed correlation between players performance and fans' emotions. Therefore, we found the direct connection between player's performance with brand's behavior on stock market.

\end{abstract}

\thispagestyle{firststyle}
\section{Introduction}

\noindent The popularity of social media website such as twitter, Facebook has increased exponentially in recent past. These social media websites have become a direct platform for the people to express their feelings, opinions about a particular topic, event or product. The proper analysis of such microblog data can help an individual or society in a better way. Therefore, the process of identification of sentiments has become a paramount in the field of data mining. It helps in determining whether a piece of text expresses a positive, negative or neutral sentiment. It can also be termed as "Opinion Mining" as the process tells the changes in public opinion about an event. Significant events like terrorist attacks, sports events, natural disasters are widely discussed topics over social media.

Twitter is one of those microbloging sites and widely used platform for emotions manifestation \& flooding the views to intended community. This assistance of twitter has turned as the habit of users. Cricket is like religion for Indians and so it can not stay away untouched  from tweeting trends. Throughout the past, it is observed that Indians are very emotionally attached with cricket. This gives us idea to capture these flowing emotions of Indian cricket lovers. From the previous studies \cite{moodpredictstockmarket}, it was observed that emotions drive the trading. Therefore, it motivates us to study one to one correspondence between the social emotions and trading behavior. 

Cricket fans follow all the liking and disliking of their role models ($Sachin~Tendulkar$). In this era of high competition, companies strive to connect with huge population of the country through their superstars as a bridge. Therefore, performance of the player who is the brand ambassador of some company becomes very crucial for advertising their brand in the market. 

In this study, we have categorized the emotions of fans for their team \& for their favourite player among 8 classes ($Joy, Fear, Anger, Anticipation, Trust,$ $Sadness, Surprise, Disgust$). In the present work, we have extracted pre-match \& post-match emotions and acquired the trading data of corresponding date. After that we have calculated the correlation between variations in emotions and trading data. 
\section{Related Work}
As per the EM hypothesis\cite{emh} changes in the stock market are driven by new information and does not follow a particular pre-defined pattern. In the past few years stock market analysis has come up with a active area of research. Various research has been done previously to extract the useful pattern in stock market and it was found that stock market does not follow a random pattern\cite{5}\cite{6} and can be analyses to find the useful patterns\cite{7}.

Recent work in field of social media data mining has showed that the microblogs like Twitter, Facebook can be used to extract the patterns of interest for various economical \& commercial changes. Further Google search queries term also play an important role to extract early indicator of spreading disease, customer opinions\cite{11}. In \cite{12}, author shows that how people emotions on the twitter can be used to predict the box office collection of a movie. 

In addition to news, there are various other factors such as people emotions and mood which plays an significant role in stock market analysis\cite{19}.

There are two approaches for  extraction of emotion analysis from the tweets/text i.e. supervised \&  unsupervised. Supervised approach requires a labeled data which is very cumbersome to obtain for a particular domain and their manual labeling is very costly and impractical. Unsupervised approach are keyword and lexicon based\cite{moodpredictstockmarket}. 	Number of domain independent lexicon are Affin, SentiWordNet, General, Inquirer, MPQA. Recently, Geo-Spatial sentiment has performed in one of the most talk event i.e \textit{'Brexit'} \cite{Geospatial},in which author shows the sentiment of users regarding the vent over the globe.   A very few studies have used the financial lexicon created by (2011). Recent work shows that domain independent lexicon are not very effective for sentiment analysis in stock market tweets\cite{oliveira2016stock}. 

In 2017, \cite{BOOK} proposed a method for analyzing the sentiment of cricket fans on Twitter. The dataset consist of tweets crawled for the 8 matches of ICC(Indian Cricket League). The results of this paper shows how the fans emotions varies frequently during the match. Further, the approach was extended by \cite{FIFAworld}, to analyses the sentiment of fans over FIFA world cup 2014 tweets. In this paper, author has used the big data approach for handling \& analyzing the tweets.

\section{Data Set Description}
In this paper our main motivation is to analyze the tweets related to Cricket matches (Champion Trophy-2017). We have collected tweets by using the twitter streaming API\footnote[1]{https://developer.twitter.com/en/docs/tutorials/consuming-streaming-data}. Tweets can be extracted in two ways either by using \textit{hashtags} or by \textit{location}. In this paper we have used hashtags \textit{('INDvsPAK', 'INDvsBAN', 'CT', 'CT'17','CT-2017', 'INDvsSA', 'INDvsSL', 'SLvsIND', 'PAKvsIND', 'BANvsIND', 'championstrophy', 'CHAMPIONTROPHY' )} for crawling the tweets.
The dataset consists of pre-match and post-match tweets for every match played by India. Total number of tweets collected were around 2 million. Statistics of match wise \textit{(pre \& post)} tweets counts are shown in table \ref{tweetstats}.
Table \ref{Count of tweets for players throughout the tournament} shows the pre \& post match tweets corresponding to each player of Indian cricket team. It clearly shows that some of the players have very less number of pre \& post tweets which implies either the player did not get chance to bat or their performance were not up to the mark.   

\begin{table}
	\centering
	\caption{Match wise Tweets statistics}
	\label{tweetstats}
	\begin{tabular}{|l|c|l|}
		\hline
		\multicolumn{1}{|c|}{\textbf{Date}} & \textbf{IND vs Opponent} & \textbf{No. of Tweets} \\ \hline
		04/06/2017                          & INDvsPAK                 & 5,34,726               \\ \hline
		08/06/2017                          & INDvsSL                  & 1,20,111               \\ \hline
		11/06/2017                          & INDvsSA                  & 2,40,112               \\ \hline
		15/06/2017                          & INDvsBAN                 & 1,35,881               \\ \hline
		18/06/2017                          & INDvsPAK                 & 5,96,809               \\ \hline
	\end{tabular}
\end{table}

In this work we have used Google trends\footnote[2]{https://trends.google.com/trends/} as a reference for our results. It is a open source web facility that determines search frequency for a particular term. The search frequency is determined relative  to the total search volume across different regions of the world. It shows the variation in the popularity of a search term with respect to time with help of graphs.  

\begin{figure}
	\centering
	\includegraphics[width=8cm,height=7cm]{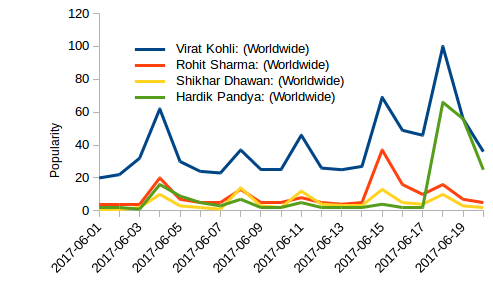}
	\caption{Google Trends: Players Popularity}
	\label{Google Trends: Players Popularity }
\end{figure}

\section{Methodology}

\begin{figure}
	\centering
	\includegraphics[width=9cm,height=7cm]{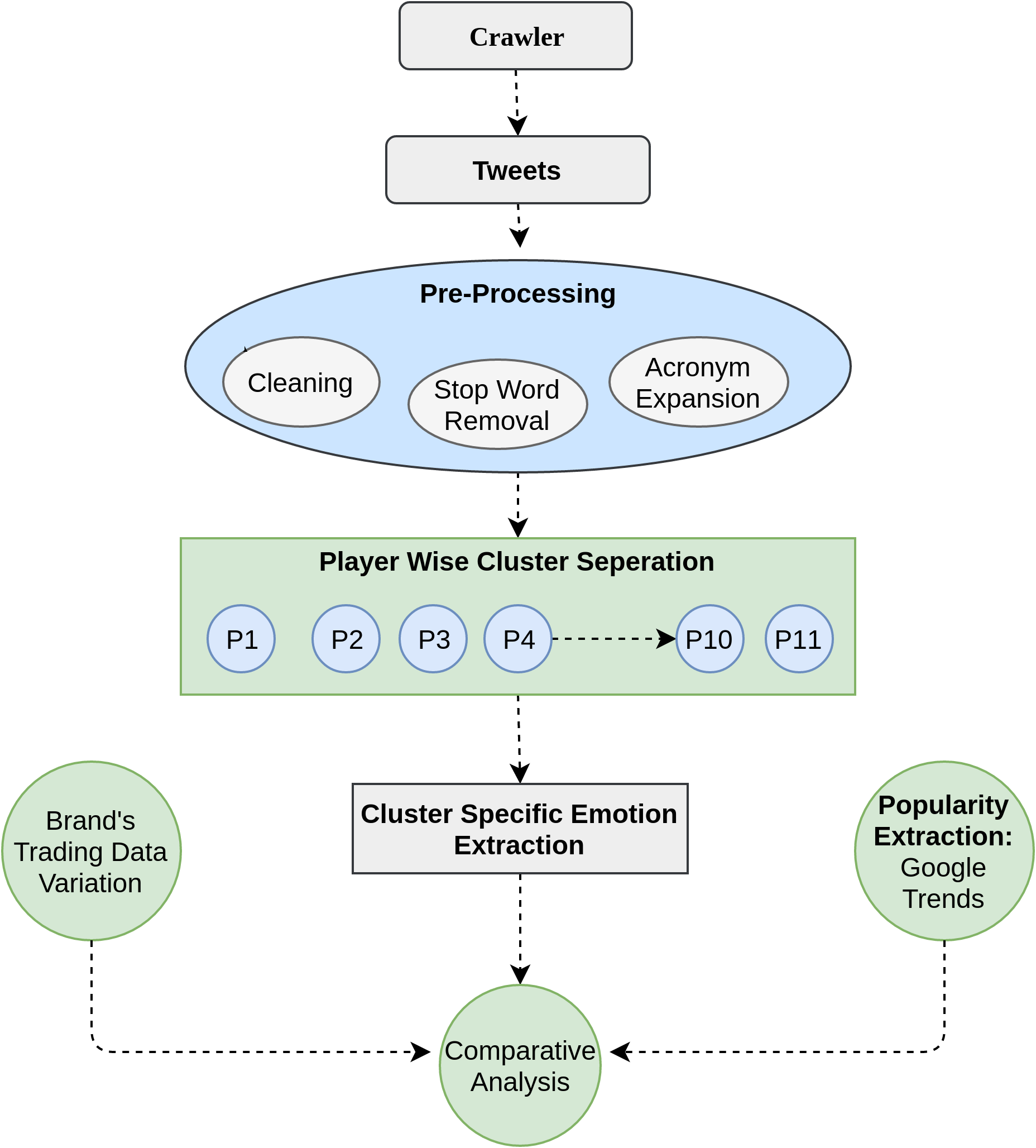}
	\caption{Architecture}
	\label{Architecture }
\end{figure}

\subsection{Data Acquisition:}
As the initial phase of data mining work, data is collected from three different sources and thereafter  correspondence is established among them.
\begin{itemize}
	\item \textbf{Match wise Tweets Crawling:} 
	We collected the streaming Tweets by using appropriate hashtags and handles corresponding to particular match. A time window is decided for pre and post match tweets  acquisition. As Tweets frequency increases when time approaches to match start time and frequency decreases down the line with time after completion. Therefore, to have the full coverage of tweets,  minimum time window is taken as 4 hours for pre and post match session. The frequency of tweets dependent on rivalry of opponents or category of match (whether it is semifinal or final). 
	
	\item \textbf{Brand identification and date specific collection of trading data:}
	Key sponsors for the team are identified, along with this each player is tagged with a brand what he endorses. All the tagged brands are taken in account and their trading\footnote[3]{http://www.bseindia.com/, https://finance.yahoo.com/lookup/} performances are acquired during the commencement of match.
	
	\item \textbf{ Player specific popularity extraction:}
	The popularity of a player can directly be decided with number of queries fired pertaining that person on Internet. The achievements of a person on global state lead others to know about the person and it's work. Popularity is a volatile property of an individual which is triggered by gets fade with time. Bringing the same concept to the context of the players the popularity of each player is extracted with the help of Google Trends.   
\end{itemize}

\subsection{Data preprocessing:}
Data accumulated from previous step are in raw form and are required to pass through filtering process so that it could attain the form suitable for model development and processing. Few steps are performed for preprocessing which are listed below.
\begin{itemize}
	\item \textbf{Data Cleaning:}
	Raw Tweets collected from the previous step contains meta data, extraneous and noisy texts which are removed in successive steps.
	\par
	Meta Data in Twitter datasets are TwitterID :$875218028102680000$, Date and Time :$ Thu~Jun~15~05:07:00 ~+0000~2017 $, location: $ Surat, ~Gujrat, ~india$, which are removed as a first step of preprocessing.  
	Extraneous texts are $@,~CT,~Opponent_1~vs~Opponent_2, ~ODI,~BCCI,$ $~CountryName$ etc. Acronyms found in the tweets are replaced by their original eg. $ICYMI:In~case~you~missed~it $, exclamatory words like$~Hooe, ~woooow, ~wowww$ are removed. Noisy texts found specific to our work were $URLs, keyword~tagging, ~links ~to ~other ~site ~etc.$. 
	\item \textbf{Stop words removal:}
	All the tweets under consideration were specifically in english language. Therefore, stopwords corresponding to  english language are removed from the sentences.
\end{itemize}

\subsection {Proposed Approach:}
In this section we have proposed an approach for emotion extraction from tweets and their impact on stock exchange.
\begin{itemize}
	\item \textbf{Tweets Clustering:}
	All the Tweets corresponding to each match under minimum pre match time window, post match time window framework are considered separately. Tweets corresponding to each match are segregated among 15 clusters, where each cluster represents a player. Player's name as hashtags, their handles eg. $imvkohli,~imvkohli,~virat$ are taken as cluster's property(keywords) for Indian Skipper Virat Kohli. These keywords are used for clubbing the new Tweets in the cluster.
	\item \textbf{Emotion Extractions and Knowledge Discovery (EEKD):}
	We have used NRC Emotion Lexicon \cite{MohammadB17wassa} \cite{MohammadB17starsem} for extracting emotions from Twitter data. NRC Emotion Lexicon consists of words and their association with 8 emotional classes (anger, fear, anticipation, trust, surprise, sadness, joy, and disgust) and two sentiments (negative and positive). All the player wise clustered tweets are segregated between pre match and post match tweets. All the tweets are tokenized to produces a bag-of-words. Each of the words are plugged-in for emotion and sentiment classification and so pre match and post match emotional score is attained for each player.     
\end{itemize}

\begin{algorithm}
	\caption{EEKD($Tweets,~Lexicon,~Brands,~S\_data$):}
	\begin{algorithmic}[1]
		\STATE $Ctweets=Pre\_Process(Tweets)$
		\FOR {Tweet in  Ctweets}
		\STATE $words=word\_tokenize(Tweet)$ 
		\FOR {word in  words}
		\STATE $\vec{Score}[word]=Emotion\_Extractor(word,Lexicon) $
		\ENDFOR
		\ENDFOR
		\FOR {Brand in  Brands}
		\STATE $\mathcal{S}[Brand]=Correlation(S\_data[brand],[\vec{Score}])$ 
		\ENDFOR
		\RETURN $[\vec{\mathcal{S}}]$ 
	\end{algorithmic} 
\end{algorithm}
\begin{table*}[]
	\centering
	\caption{Count of tweets for players throughout the tournament}
	\label{Count of tweets for players throughout the tournament}
	\begin{tabular}{|c|c|c|c|c|c|c|}
		\hline
		\textbf{Name}                & \multicolumn{1}{l|}{\textbf{Tweets}}               & \multicolumn{1}{l|}{\textbf{INDvsPak}}               & \multicolumn{1}{l|}{\textbf{INDvsSA}}                & \multicolumn{1}{l|}{\textbf{INDvsBAN}}               &
		\multicolumn{1}{l|}{\textbf{INDvsPAK}}               & \multicolumn{1}{l|}{\textbf{INDvsSL}}               \\ \hline
		\textbf{Virat Kohli(VK)}         & \begin{tabular}[c]{@{}c@{}}Pre\\ Post\end{tabular} & \begin{tabular}[c]{@{}c@{}}8004\\ 32830\end{tabular} & \begin{tabular}[c]{@{}c@{}}2537\\ 10742\end{tabular} & \begin{tabular}[c]{@{}c@{}}1619\\ 12001\end{tabular} & \begin{tabular}[c]{@{}c@{}}5550\\ 12033\end{tabular} &
		\begin{tabular}[c]{@{}c@{}}2902\\ 11753\end{tabular} \\ \hline
		\textbf{Rohit Sharma(RS)}        & \begin{tabular}[c]{@{}c@{}}Pre\\ Post\end{tabular} & \begin{tabular}[c]{@{}c@{}}1842\\ 24790\end{tabular} & \begin{tabular}[c]{@{}c@{}}439\\ 1662\end{tabular}   & \begin{tabular}[c]{@{}c@{}}1386\\ 7186\end{tabular}  & \begin{tabular}[c]{@{}c@{}}766\\ 3451\end{tabular}   &
		\begin{tabular}[c]{@{}c@{}}1157\\ 9285\end{tabular}\\ \hline
		\textbf{Shikhar Dhawan(SD)}      & \begin{tabular}[c]{@{}c@{}}Pre\\ Post\end{tabular} & \begin{tabular}[c]{@{}c@{}}1203\\ 10080\end{tabular} & \begin{tabular}[c]{@{}c@{}}1331\\ 9624\end{tabular}  & \begin{tabular}[c]{@{}c@{}}1451\\ 3654\end{tabular}  & \begin{tabular}[c]{@{}c@{}}1618\\ 3770\end{tabular}  &
		\begin{tabular}[c]{@{}c@{}}3027\\ 2453\end{tabular}\\ \hline
		\textbf{Yuvraj Singh(YS)}        & \begin{tabular}[c]{@{}c@{}}Pre\\ Post\end{tabular} & \begin{tabular}[c]{@{}c@{}}1145\\ 10816\end{tabular} & \begin{tabular}[c]{@{}c@{}}670\\ 1234\end{tabular}   & \begin{tabular}[c]{@{}c@{}}509\\ 1106\end{tabular}   & \begin{tabular}[c]{@{}c@{}}240\\ 880\end{tabular}    &
		\begin{tabular}[c]{@{}c@{}}267\\ 126\end{tabular}\\ \hline
		\textbf{Kedar Jadhav(KJ)}        & \begin{tabular}[c]{@{}c@{}}Pre\\ Post\end{tabular} & \begin{tabular}[c]{@{}c@{}}580\\ 981\end{tabular}    & \begin{tabular}[c]{@{}c@{}}426\\ 1088\end{tabular}   & \begin{tabular}[c]{@{}c@{}}476\\ 1909\end{tabular}   & \begin{tabular}[c]{@{}c@{}}245\\ 4338\end{tabular}   &
		\begin{tabular}[c]{@{}c@{}}864\\ 479\end{tabular}\\ \hline
		\textbf{MS Dhoni(MSD)}            & \begin{tabular}[c]{@{}c@{}}Pre\\ Post\end{tabular} & \begin{tabular}[c]{@{}c@{}}6453\\ 9100\end{tabular}  & \begin{tabular}[c]{@{}c@{}}1565\\ 5001\end{tabular}  & \begin{tabular}[c]{@{}c@{}}1541\\ 1602\end{tabular}  & \begin{tabular}[c]{@{}c@{}}2135\\ 6245\end{tabular}  &
		\begin{tabular}[c]{@{}c@{}}7270\\ 2412\end{tabular}\\ \hline
		\textbf{Hardik Pandya(HP)}       & \begin{tabular}[c]{@{}c@{}}Pre\\ Post\end{tabular} & \begin{tabular}[c]{@{}c@{}}755\\ 3788\end{tabular}   & \begin{tabular}[c]{@{}c@{}}431\\ 1028\end{tabular}   & \begin{tabular}[c]{@{}c@{}}331\\ 820\end{tabular}    & \begin{tabular}[c]{@{}c@{}}1276\\ 14303\end{tabular} &
		\begin{tabular}[c]{@{}c@{}}1235\\ 269\end{tabular}\\ \hline
		\textbf{Ravindra Jadeja(RJ)}     & \begin{tabular}[c]{@{}c@{}}Pre\\ Post\end{tabular} & \begin{tabular}[c]{@{}c@{}}708\\ 9106\end{tabular}   & \begin{tabular}[c]{@{}c@{}}322\\ 503\end{tabular}    & \begin{tabular}[c]{@{}c@{}}230\\ 1217\end{tabular}   & \begin{tabular}[c]{@{}c@{}}238\\ 7908\end{tabular}   &
		\begin{tabular}[c]{@{}c@{}}1185\\ 352\end{tabular}\\ \hline
		\textbf{Bhuvneshwar Kumar(BK)}   & \begin{tabular}[c]{@{}c@{}}Pre\\ Post\end{tabular} & \begin{tabular}[c]{@{}c@{}}566\\ 1453\end{tabular}   & \begin{tabular}[c]{@{}c@{}}350\\ 810\end{tabular}    & \begin{tabular}[c]{@{}c@{}}414\\ 1318\end{tabular}   & \begin{tabular}[c]{@{}c@{}}382\\ 2365\end{tabular}   &
		\begin{tabular}[c]{@{}c@{}}1163\\ 248\end{tabular}\\ \hline
		\textbf{Ravichandran Ashwin(RA)} & \begin{tabular}[c]{@{}c@{}}Pre\\ Post\end{tabular} & \begin{tabular}[c]{@{}c@{}}405\\ 415\end{tabular}    & \begin{tabular}[c]{@{}c@{}}1020\\ 1393\end{tabular}  & \begin{tabular}[c]{@{}c@{}}237\\ 1261\end{tabular}   & \begin{tabular}[c]{@{}c@{}}271\\ 3336\end{tabular}   &
		\begin{tabular}[c]{@{}c@{}}268\\ 944\end{tabular}\\ \hline
		\textbf{Jasprit Bumrah(JB)}      & \begin{tabular}[c]{@{}c@{}}Pre\\ Post\end{tabular} & \begin{tabular}[c]{@{}c@{}}707\\ 563\end{tabular}    & \begin{tabular}[c]{@{}c@{}}1316\\ 1777\end{tabular}  & \begin{tabular}[c]{@{}c@{}}540\\ 1250\end{tabular}   & \begin{tabular}[c]{@{}c@{}}270\\ 5380\end{tabular}   &
		\begin{tabular}[c]{@{}c@{}}623\\ 221\end{tabular}\\ \hline
	\end{tabular}
\end{table*} 
Algorithm [1] explains the successive steps to be performed for knowledge discovery. Algorithm takes Tweets corresponding to particular player, Emotion Lexicon as $Lexicon$, list of Brands endorsed by the player as $Brands$, and trading performance of the brand on stock market as $S\_data$. In line-1  preprocessed tweets are received as $Ctweets$ which are further processed for tokenized and plugged-in to $Emotion\_Extractor$ along with emotion Lexicon.
\begin{equation}
 r(X,Y) = \frac{N\sum{XY}-(\sum{X}\sum{Y})}{\sqrt{ [N \sum{x^2}-(\sum{x})^2 ][N \sum{y^2}-(\sum{y})^2 }]} \label{eq}
\end{equation}
 
This function finds the availability of category wise emotion index in each of the words and keeps increasing the score as new words are encountered. $\vec{Score}$ is the vector of 10 in which each index represents particular emotion or sentiment.  
Score Vector $\vec{Score}$ is calculated for all the words in tweet dataset. In line 8,9 correlation is calculated between each brand endorsed and emotion score $\vec{Score}$ using equation \ref{eq}. Correlation score represented by $\vec{\mathcal{S}}$ is the extracted knowledge as an output of the given algorithm.



\subsection{Results and Discussion}
The proposed architecture extracts emotions and sentiments from a cricket loving country and shows how do emotions affect stock market's performance of various commercial brands whose brand ambassadors actively participate in the game. The study reveals that performance of the brand ambassador has direct impact on the brand's price on stock exchange. 
\par

Tables [\ref{Emotion Analysis of Followers for Batsmen},\ref{Emotion Analysis of Followers for Bowlers}] show the relative strength of various emotions of Indian cricket fans when their team \lq India' plays with different countries. All 5 cricket matches of an ICC event (ICC Champions Trophy) played by their team against \lq Pakistan\rq(league match),\lq Srilanka\rq(league match), \lq SouthAfrica\rq(league match), \lq Bangladesh \rq(Semifinal) and final again with \lq Pakistan' are taken under consideration for analysis. The results show how the fans' emotions change as the game commences.  Because of uneven counting of emotions for each player in particular emotional category, in this work all the emotions are scaled down between 0 to 1 with respect to total number of emotions, corresponding to each player. Figure \ref{Google Trends: Players Popularity } shows the variance of particular emotion with matches and popularity of the players can be seen through Google trend over the globe . The rows in the table \ref{Emotion Analysis of Followers for Batsmen} represent the values of pre and post match emotions respectively. The key players like $Virat~Kohli,~Rohit ~Sharma, ~Shikhar ~Dhawan$ captured the largest part of Tweets and so the emotions. $Yuvraj~Singh$, an experienced player who was striving to comeback through this series in the playing eleven after a long time has huge burden  of expectations . As he could not performed up to the mark of fans therefore $sadness$ emotions gets increased in every match played by him. $Hardik~Pandiya$ performed as a dark horse in the final. So he could reinforce all the positive emotions $Trust,~ joy$ among his fans and could succeed to suppress negative emotions like $ Fear,~Anger,~Disguest$.
\par
Few emotions are found very consistent throughout the matches and are noticed as facing the direct impact of whether team has won or lost. We observed that changes in emotions between pre and post match sessions were positive when that individual player performed well and negative otherwise. 
Along with these emotions we have collected sentiments through pre match session tweets and post match session tweets with binary values as positive or negative. It is observed that even if player could not perform well but if team wins, sentiments are noted as positive. This shows that sentiments are more consistent with the Team than with the player. 
\par
Trading price of commercial brands are found to have transitive relationship with their brand ambassador's performance in the match as shown in table [\ref{Rohit Sharma}, \ref{Virat Kohli}, \ref{Dhawan}]. Good performance spreads happiness among fans that gives jump in trading price and bad performance discourages the fans and their commercial activities so their average mood leads to fall in stock prices. 
\par 
First column in table [\ref{Emotion Analysis of Followers for Batsmen}, \ref{Emotion Analysis of Followers for Bowlers}] are the names of cricketers andsecond column denotes sequence of the match along with players performance, all-rounder's performance are shown as $bat\_score/bowl\_score$ respectively. Name acronyms in table [\ref{Emotion Analysis of Followers for Batsmen}, \ref{Emotion Analysis of Followers for Bowlers}] can be referred from table \ref{Count of tweets for players throughout the tournament}.

\begin{figure*}
	\centering
	\subfigure[Batsmen]{\label{df}\includegraphics[width=7cm,height=5cm]{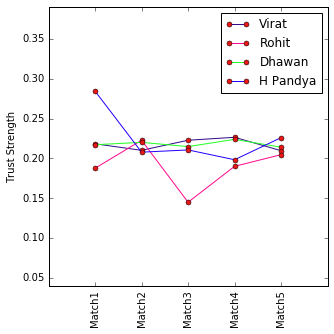}}
	\subfigure[Bowlers]{\label{df} \includegraphics[width=7cm,height=5cm]{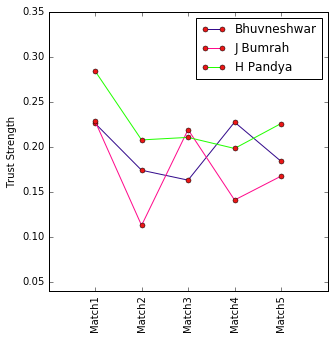}}
	\caption{Variation of trust in a tournament}
	\label{trust}
\end{figure*}

\subsection{Conclusion}
The high popularity of cricket in Asia with respect to all other part of the world stimulated us to know the emotions of cricket lovers. The presence od strong emotions of cricket fans for their favorite player/ team over Tweets motivated us to extract the emotions/ sentiments and to witness its impact in financial domain. 
\par In this work we have collected tweets corresponding to ICC-CT2017(Champions Trophy) held in England in month of June. We have collected all the tweets pertaining Indian team through out the tournament. For all the matches we have extracted emotions/sentiments before and after the match. We have established one to one correspondence between players' performance and fan's emotions fro that player and this study uses total 8 category of emotions and 2 sentiments for this purpose. This work picks few key players of the team on performance basis and finds their commercial endorsements.
Next, we analyzed the impact of their performance on the emotions of the twitter followers as well as on the stock prices of the brands which they are endorsing. As a reference we have used Google Trends to verify the influence of players performance all over the globe
\par 
In our observations we have noticed high correlation of  player's performance with corresponding emotions/ sentiments as well as on stock prices of the brand they endorsed.

\begin{table*}[]
	\centering
	\caption{Emotion Analysis of Followers for Batsmen with Team performance as: WLWWL}
	\label{Emotion Analysis of Followers for Batsmen}

\end{table*}

\begin{table*}[]
	\centering
	\caption{Emotion Analysis of Followers for Bowlers with Team performance as: WLWWL}
	\label{Emotion Analysis of Followers for Bowlers}
	%
\end{table*}

\begin{table*}[]
	\centering
	\caption{Virat Kohli: Trust and Business}
	\label{Virat Kohli}
	\begin{tabular}{|l|l|l|l|l|l|l|}
		\hline
		&      & \textbf{Match-1} & \textbf{Match-2} & \textbf{Match-3} & \textbf{Match-4} & \textbf{Match-5} \\ \hline
		\multirow{2}{*}{\textbf{Trust}}   & Pre  & 0.1880           & 0.2277           & 0.2070           & 0.2104           & 0.2153           \\ \cline{2-7} 
		& Post & .2486            & 0.1925           & 0.2385           & 0.2425           & 0.2040           \\  \hline
		\hline
		
		\multirow{2}{*}{\textbf{MRF}}     & Pre  & 67716            & 68986            & 72901.20313      & 71102.10156      & 71978.10156      \\ \cline{2-7} 
		& Post & 68820.5          & 72901.20313      & 73331.60156      & 71978.10156      & 71760.20313      \\ \hline
		\multirow{2}{*}{\textbf{Pepsico}} & Pre  & 117.669998       & 117.050003       & 115.940002       & 117.370003       & 116.860001       \\ \cline{2-7} 
		& Post & 117.699997       & 115.940002       & 116.230003       & 116.860001       & 117.290001       \\ \hline
		\multirow{2}{*}{\textbf{PNB}}     & Pre  & 153.050003       & 152.350006       & 152.149994       & 152              & 147.649994       \\ \cline{2-7} 
		& Post & 149.5            & 152.149994       & 148.399994       & 147.649994       & 147.25           \\ \hline
		\multirow{2}{*}{\textbf{Audi}}    & Pre  & 675.5            & 685.099976       & 682              & 705              & 711.349976       \\ \cline{2-7} 
		& Post & 680              & 682              & 707              & 711.349976       & 714.599976       \\ \hline
	\end{tabular}
\end{table*}


\begin{table*}[]
	\centering
	\caption{Rohit Sharma: Trust and Business}
	\label{Rohit Sharma}
	\begin{tabular}{|l|l|l|l|l|l|l|}
		\hline
		&      & \textbf{Match-1} & \textbf{Match-2} & \textbf{Match-3} & \textbf{Match-4} & \textbf{Match-5} \\ \hline
		\textbf{Trust}                   & Pre  & 0.1607           & 0.2403           & 0.1182           & 0.1735           & 0.2093           \\ \hline
		\textbf{}                        & Post & 0.2146           & 0.2051           & 0.1718           & 0.2066           & 0.2              \\ \hline \hline
		\multirow{2}{*}{\textbf{CEAT}}   & Pre  & 1860.404297      & 1883.064209      & 1896.23291       & 1846.987305      & 1849.869507      \\ \cline{2-7} 
		& Post & 1880.828125      & 1896.23291       & 1891.114624      & 1849.869507      & 1842.564453      \\ \hline
		\multirow{2}{*}{\textbf{Maggie}} & Pre  & 6714.200195      & 6669.799805      & 6667.5           & 6684.350098      & 6759.75          \\ \cline{2-7} 
		& Post & 6677.200195      & 6667.5           & 6660.149902      & 6759.75          & 6738.350098      \\ \hline
	\end{tabular}
\end{table*}

\begin{table*}[]
	\centering
	\caption{Dhawan: Trust and Business}
	\label{Dhawan}
	\begin{tabular}{|l|l|l|l|l|l|l|}
		\hline
		&      & \textbf{Match-1} & \textbf{Match-2} & \textbf{Match-3} & \textbf{Match-4} & \textbf{Match-5} \\ \hline
		\multirow{2}{*}{\textbf{Trust}}      & Pre  & 0.2132           & 0.2101           & 0.2072           & 0.1696           & 0.2363           \\ \cline{2-7} 
		& Post & 0.2211           & 0.2301           & 0.2224           & 0.2781           & 0.1917           \\ \hline \hline
		\multirow{2}{*}{\textbf{Canara}}     & Pre  & 360.049988       & 362.700012       & ,360.450012      & 355.350006       & 352.049988       \\ \cline{2-7} 
		& Post & 362.649994       & 360.450012       & 346              & 352.049988       & 355.450012       \\ \hline
		\multirow{2}{*}{\textbf{Big Bazaar}} & Pre  & 329.899994       & 382.600006       & 393.350006       & 388.549988       & 373.700012       \\ \cline{2-7} 
		& Post & 346.399994       & 393.350006       & 394.649994       & 373.700012       & 393.149994       \\ \hline
	\end{tabular}
\end{table*}

\begin{table*}[]
	\centering
	\caption{Sentiments Analysis with performance of Team}
	\label{my-label}
	%
 \\ \hline
	\end{tabular}
\end{table*}




\end{document}